\documentclass[twocolumn]{aastex62}

\usepackage{amssymb}
\usepackage{amsmath}
\usepackage{amssymb}
\usepackage{empheq}
\usepackage{mathrsfs}

\usepackage{etoolbox}
\usepackage{lmodern}

\newcommand{\G}{\mathcal{G}} 
\newcommand{\Ham}{\mathcal{H}} 
\newcommand{\M}{M_{\star}} 
\newcommand{\tauap}{\tau_{m}'} 
\newcommand{\taua}{\tau_{m}} 
\newcommand{\taub}{\bar{\tau}_{m}} 
\newcommand{\tauep}{\tau_{e}'} 
\newcommand{\taue}{\tau_{e}} 
\newcommand{\K}{\mathcal{K}} 
\newcommand{\twave}{\tau_{\rm{wave}}}
\newcommand{\ares}{\alpha_{\rm{res}}}

\newcommand{\deriv}[2]{\frac{\mathrm{d} #1}{\mathrm{d} #2}}

\begin{document}

\title{Dissipative Capture of Planets Into First-Order Mean-Motion Resonances}

\author{Konstantin Batygin}
\affiliation{Division of Geological and Planetary Sciences California Institute of Technology, Pasadena, CA 91125, USA}

\author{Antoine C. Petit}
\affiliation{Laboratoire Lagrange, Universit\'e Cote d'Azur, CNRS, Observatoire de la Cote d'Azur, Nice, France}


\begin{abstract}
The emergence of orbital resonances among planets is a natural consequence of the early dynamical evolution of planetary systems. While it is well-established that convergent migration is necessary for mean-motion commensurabilities to emerge, recent numerical experiments have shown that the existing adiabatic theory of resonant capture provides an incomplete description of the relevant physics, leading to an erroneous mass scaling in the regime of strong dissipation. In this work, we develop a new model for resonance capture that self-consistently accounts for migration and circularization of planetary orbits, and derive an analytic criterion based upon stability analysis that describes the conditions necessary for the formation of mean-motion resonances. We subsequently test our results against numerical simulations and find satisfactory agreement. Our results elucidate the critical role played by adiabaticity and resonant stability in shaping the orbital architectures of planetary systems during the nebular epoch, and provide a valuable tool for understanding their primordial dynamical evolution.
\end{abstract}


\section{Introduction}

Orbital resonances facilitate long-term exchange of energy and angular momentum within planetary systems, thereby playing a critical role in their long-term evolution. The preference for orbital commensurability -- first quantified in a statistically rigorous manner by \citet{Dermott1968a,Dermott1968b} -- is a well-known attribute of the solar system's architecture, which is particularly pronounced among the satellites of Jupiter, Saturn, and Uranus (see \citealt{MD99} and the references therein). Beyond the realm of the solar system, resonant configurations can be found in appreciable proportion within the census of giant and sub-Jovian exoplanets alike (e.g., \citealt{2016MNRAS.455L.104G,2016Natur.533..509M,2017NatAs...1E.129L,2020MNRAS.496.3101P,2022ApJ...925...38N,2023AJ....165...33D}). Intriguingly, the importance of resonant dynamics likely goes well beyond the population of planetary systems that are presently entrained in mean-motion commensurabilities. That is to say, the evolutionary role played by \textit{transient} mean-motion resonances is almost certainly more significant than a superficial examination of the data may indicate. To this end, numerous lines of evidence suggest that the outer solar system itself originated in a compact multi-resonant configuration before becoming temporarily unstable and eventually settling in its current state by way of dynamical friction \citep{2010ApJ...716.1323B,2012AJ....144..117N}. Such a sequence of events may in fact constitute a relatively typical post-nebular evolutionary path of planetary systems, and recent modeling has shown that both the period ratio distribution, as well as the degree of intra-system uniformity of short-period super-Earths, can be satisfactorily reproduced if the majority of systems originate as resonant chains that subsequently relax towards more widely-spaced orbits through dynamical instabilities \citep{Izidoro2017,Izidoro2021,GB22a,BM23}.

Despite their extant and inferred prevalence, mean motion resonances do not arise as an innate byproduct of the planet formation process itself. Instead, they are established as a consequence of orbital convergence facilitated by dissipative effects \citep{Goldreich1965,1982CeMec..27....3H}. Within protoplanetary nebulae, this occurs naturally due to planet-disk interactions (i.e., type-I migration; \citealt{1980ApJ...241..425G,1997Icar..126..261W}) -- particularly in the inner regions of disks, where magnetospheric cavities create bonafide traps for planetary orbits \citep{2006ApJ...642..478M}.

In addition to the necessity of convergent migration, resonance capture requires stability of the resonant equilibrium and adiabaticity. Crudely speaking, this means that dissipative torques must not exceed gravitational perturbations in magnitude, and that resonant dynamics must operate ``faster" than the timescale associated with extrinsic (that is, disk-driven) forcing of the orbits. In this vein, \citet{B15} proposed an analytic criterion for adiabatic capture in the unrestricted 3-body problem, by equating the resonant libration (bounded oscillation) period to the migratory resonance-crossing time. While this criterion yields quantitatively adequate results in the regime where orbital migration ensues in absence of other dissipative effects, the recent simulation suite of \citet{K23} has shown that the behavior of resonance capture is qualitatively different if convergent migration is accompanied by strong eccentricity damping. Evidently, disk-driven orbital circularization alters the efficiency of resonance capture in a non-trivial manner.

The principle goal of this Letter is to understand the process of resonance capture in presence of direct dissipation, from theoretical grounds. That is, in this work, we employ perturbation theory to quantify conditions under which stable resonant dynamics can be established, derive an analytic criterion for such dissipative capture, and confirm our results with numerical experiments. The remainder of the manuscript is organized as follows. In section \ref{sec:section2}, we outline a simplified sketch of the resonance stability argument within the context of the circular restricted 3-body problem. We generalize our analytical framework to the unrestricted elliptic problem and compare our results with numerical simulations in section \ref{sec:section3}. We summarize and discuss our findings in section \ref{sec:section4}.

\section{The Restricted Problem} \label{sec:section2}

As the simplest starting point for our analysis, we adopt the circular restricted 3-body problem as a paradigm, wherein one of the secondary bodies is taken to have negligible mass, while the other is assumed to reside on a circular orbit. A similar approach has been undertaken in the recent study of \citet{2023arXiv230203070H}. To be clear, we make the restricted approximation in this section strictly for comprehensibility: the work of \citet{1984CeMec..32..307S,1986CeMec..38..175W}, as well as a number of more recent studies \citep{BM13,2017A&A...607A..35P,2019AJ....158..238H} have shown how the perturbative treatment of first-order resonances within the restricted problem can be generalized to the full 3-body problem\footnote{Within the context of the full 3-body problem, results depend predominantly on the sum of the planetary masses, rather than their ratio \citep{2013ApJ...774..129D,Deck15}}, and we carry out this generalization in the next section.

\bigskip
\bigskip

\subsection{Perturbation Theory}

\paragraph{Model Hamiltonian.} Upon averaging over short-periodic terms and expanding the interaction potential (i.e., the disturbing function) to leading order in eccentricity and inclination, the governing Hamiltonian for a $k:k-1$ mean-motion resonance takes the form:
\begin{align}
\label{Hammy}
\Ham &= - \frac{\G\,\M}{2\,a'}-\frac{\G\,m}{a'} \nonumber \\ 
&\times\,f\,e' \cos(k\,\lambda'-(k-1)\,n\,t-\varpi').
\end{align}
In the above expression, the Keplerian orbital elements have their usual meanings, $f$ is constant of order unity\footnote{The coefficient $f$ depends only on the semi-major axis ratio, and evaluates to $f\approx1.2$ for $k=2$ and $f\approx0.8\,k$ for $k\geqslant 3$.}, and the primed variables refer to the outer body, which we take to be massless. This choice circumvents any consideration of over-stable librations, which can ensue if the massive perturber resides on an exterior orbit \citep{Deck15}. We note however, that the statistical analysis of \citet{2023arXiv230203070H} indicates that even under a reversed mass-ordering, over-stable librations are expected to be rare in real protoplanetary disks. In addition, we have explicitly written the mean longitude of the inner orbit explicitly as a product of its mean motion, $n = \sqrt{\G\,\M/a^3}$, and time. In other words, the physical setup of our problem has: $M\gg m \neq 0; m'=0; e=0; \lambda = n\,t$.

To simplify the functional form of $\mathcal{H}$, we follow the well-documented procedure of expanding the leading (Keplerian) term of equation (\ref{Hammy}) in the vicinity of exact commensurability, to second order in $\delta \mathcal{L} = \mathcal{L} - [\mathcal{L}]$, where $\mathcal{L}=\sqrt{\G\,\M\,a’}$ and $[\mathcal{L}] = \sqrt{\G\,\M\,[a’]}=\sqrt{\G\,\M\,(k/(k-1))^{2/3}\,a}$ represents the maximal specific angular momentum of the test-particle orbit, evaluated at the nominal resonance semi-major axis, $[a']$. Switching to the canonically conjugated action-angle variables (e.g., \citealt{Peale1986})
\begin{align}
&\Phi = [\mathcal{L}]\,(e')^2/2 &\phi = k\,\lambda'-(k-1)\,n\,t-\varpi' \nonumber \\
&\Psi = \delta\mathcal{L} - k\, \Phi &\psi = \lambda',
\label{vars}
\end{align}
Hamiltonian (\ref{Hammy}) takes the familiar form of the second fundamental model for resonance \citep{1983CeMec..30..197H}:
\begin{align}
\Ham &= n' (k\,\Phi+\Psi)-n\,(k-1)\,\Phi  \nonumber \\
& -\frac{3}{2}\frac{n'}{[\mathcal{L}]}(k\Phi+\Psi)^2-\mathcal{F}\,\sqrt{2\,\Phi}\cos(\phi),
\label{SecondHammy}
\end{align}
where $\mathcal{F} = (\G\,m/[a'])\,(f/\sqrt{[\mathcal{L}]})$ is a constant. Because the angle $\psi$ does not appear within $\Ham$, the evolution of the conjugated action, $\Psi$, is dictated entirely by extrinsic forces.

\paragraph{The case of pure migration.} Before proceeding to consider the full dissipative problem, let us pause and recall some qualitative aspects of the well-studied instance of pure orbital migration, where external (non-Hamiltonian) forces do not affect the $(\Phi,\phi)$ degree of freedom directly. In this case, it is easy to see that convergent migration will cause the action $\Psi$ to diminish without bound, changing the topological structure of the phase-space portrait of $\Ham$ in concert \citep{1983CeMec..30..197H,B15}. If the evolution of $\Psi$ is slow compared to resonant dynamics, then an adiabatic invariant -- which corresponds to the phase-space area encircled by the orbit -- emerges as a quasi-integral of motion \citep{1982CeMec..27....3H,1984PriMM..48..197N}. 

In the practically important case where orbits originate with zero eccentricity far away from resonance, the initially occupied phase-space area is null, meaning that as long as the adiabatic condition is satisfied\footnote{An important additional caveat is that no encounters with the separatrix take place.}, the trajectory must remain confined to the $\phi=\pi$ resonant equilibrium point (since it encapsulates zero phase-space area). Simultaneously, as $\Psi$ evolves to highly negative values, the eccentricity grows perpetually as $e'\sim \sqrt{-2\,\Psi/(k\,[\mathcal{L}])}$. In this manner, forces that facilitate the convergent migration of orbits translate to sustained eccentricity excitation, once the resonant lock is established.

\paragraph{The case of concurrent migration and circularization.} The arguably more physically realistic scenario -- wherein convergent migration occurs together with efficient orbital damping -- is different from the aforementioned case of pure migration in a number of important ways. First and foremost, the evolution of the action $\Psi$ is no longer unbounded, and instead stabilizes at an equilibrium value, which in turn dictates the equilibrium eccentricity. To quantify this, consider the following generic parameterizations of semi-major axis decay and orbital circularization:
\begin{align}
&\frac{1}{a'}\frac{da'}{dt} =- \frac{1}{\tauap}- \frac{2\,e'^2}{\tauap/  \mathcal{K} } &\frac{1}{e'}\frac{de'}{dt} = -\frac{\mathcal{K} }{\tauap},
\label{taua}
\end{align}
where $\tauap$ is the convergent migration timescale and $\mathcal{K} = \tauap/\tau_e'$ is the ratio of semi-major axis and eccentricity damping timescales.

It is worth noting that \citep{2022arXiv221203608P} have recently shown that the eccentric contribution to semi-major axis damping in equations (\ref{taua}) arises self-consistently within planet-disk interaction formulae that are routinely implemented in $N-$body codes to mimic the effects of the gaseous nebula \citep{PL2000}. Moreover, for type-I migration, $\mathcal{K}$ has a well-defined dependence on the the geometric aspect ratio, $h/r$, \citep{Tanaka2002,Tanaka2004}:
\begin{align}
\K = \bigg(\frac{1}{e'}\frac{de'}{dt} \bigg) \bigg(\frac{1}{a'}\frac{da'}{dt} \bigg)^{-1} \propto \bigg(\frac{h}{r} \bigg)^{-2},
\label{Kdef}
\end{align}
though the dimensionless pre-factor of this dependence is specified by the disk's particular structure. As an example, for a locally isothermal \citet{1963MNRAS.126..553M} type disk (where the surface density varies inversely with the semi-major axis -- i.e., $\Sigma \propto 1/a'$), this pre-factor is approximately $0.2$, such that $\K \sim \mathcal{O}(10^2)$ for $h/r \sim 0.05$ (see also \citealt{2002ApJ...567..596L}). Nevertheless, values for $\K$ that are substantially higher (and lower) are expected to arise in realistic model nebulae; we will revisit the relevant scalings in the next section.

With the relevant damping formulae defined, it is straightforward to derive the equilibrium value for $e'$. Recalling the definition of $\Psi$ from equations (\ref{vars}), let us set the time-derivative of $\Psi$ equal to zero:
\begin{align}
\frac{d\Psi}{dt} &= \frac{1}{2}\sqrt{\frac{\G\,\M}{a'}}\frac{d a'}{dt} - 2\, [\mathcal{L}]\, e' \,\frac{de'}{dt} \nonumber \\
&= - \frac{\mathcal{L}}{2}\frac{1+4\,\K\,\Phi/[\mathcal{L}] }{\tauap} + 2\,\K \,\frac{k\,\Phi}{\tauap} = 0.
\label{Psieq}
\end{align}
It is expected that equilibrium will be reached close to nominal resonance\footnote{Notice that this is not the case for the case of dissipative divergent migration, where the system can follow equilibrium loci far away from nominal commensurability \citep{2019A&A...625A...7P,2021AJ....162...16G}.} such that $\delta\mathcal{L}\approx0$. Thus, replacing $\mathcal{L}$ by $[\mathcal{L}]$ in the above equation and recalling from equations (\ref{vars}) that $\Phi=[\mathcal{L}]\,e'^2/2$, we obtain:
\begin{align}
\big(e'\big)_{\rm{eq}}\rightarrow\frac{1}{\sqrt{2\,(k-1)\,\K}} \sim \frac{h}{r}.
\label{eeq}
\end{align}
Indeed, resonantly-excited planetary eccentricities within protoplanetary nebulae are expected to be comparable to the disk aspect ratio (which is itself equal to the inverse Mach number of the Keplerian flow), as the above expression suggests (e.g., see also \citealt{2018CeMDA.130...54P}).

The terminal step of the calculation is to evaluate the stability of the resonant fixed point. Here, a second important distinction with the pure migration case comes into view: in presence of explicit eccentricity damping, the phase-space portrait rotates counter-clockwise, such that the equilibrium value of the critical angle, $\phi$, shifts to $\phi_{\rm{eq}}=\pi+\epsilon$, where $\epsilon$ is determined by the strength of dissipative effects \citep{BM13diss}. Importantly, criticality is achieved when $\epsilon\rightarrow \pi/2$ and $\delta\mathcal{L}\rightarrow0$ -- a configuration where the resonant torque is maximized. It thus follows that in this state,
\begin{align}
\big(\Psi)_{\rm{eq}}\rightarrow - k \, \big( \Phi \big)_{\rm{eq}} = \frac{[\mathcal{L}]}{4\,(k-1)\,\K}.
\label{Psieq}
\end{align}

\begin{figure}[t!]
\centering
\includegraphics[width=\columnwidth]{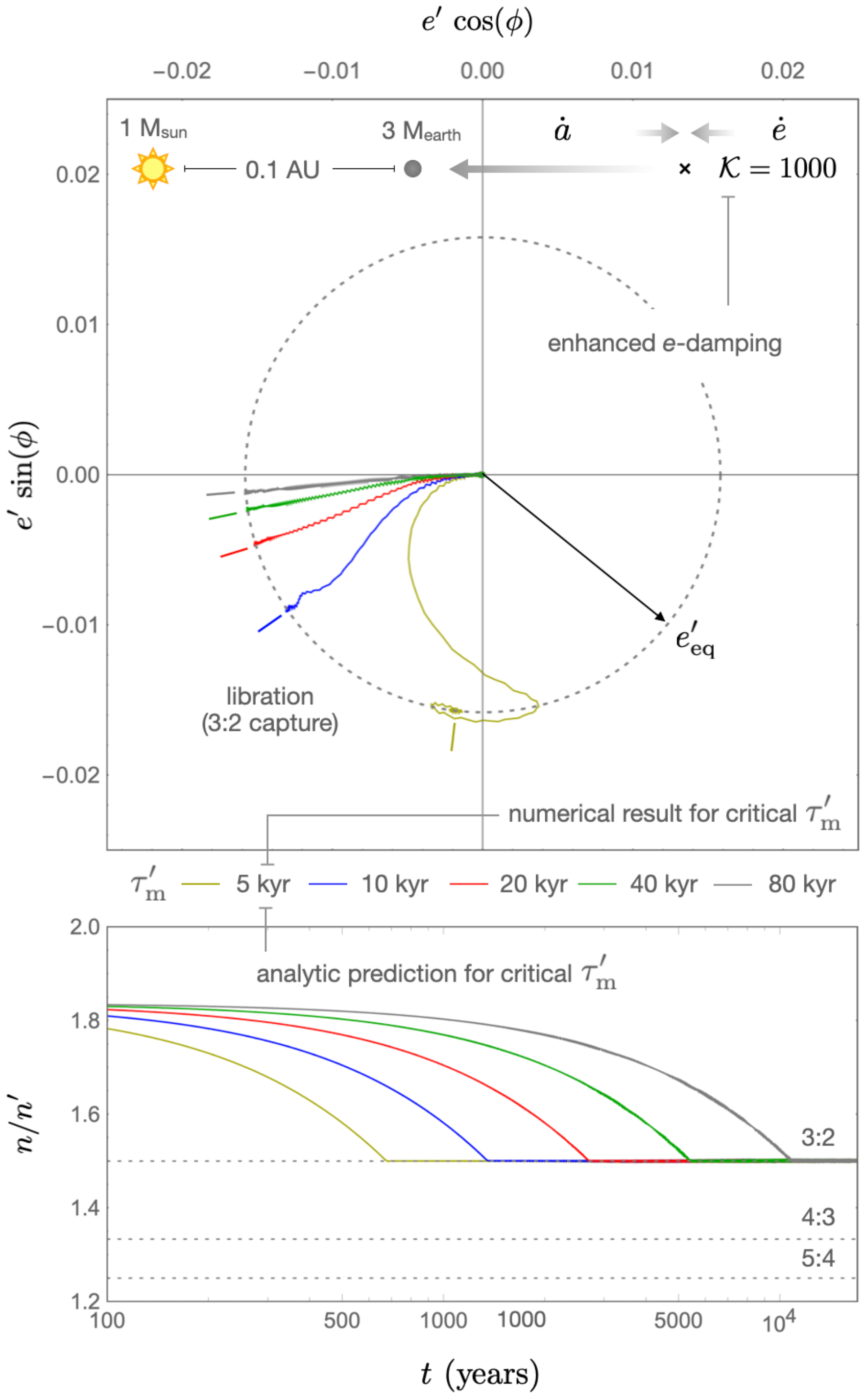}
\caption{Numerical simulations of a $k=3$ resonant encounter within the context of the circular restricted 3-body problem. The top and bottom panels depict phase-space evolution and the time-series of the orbital frequency ratio, respectively. Differently colored curves correspond to distinct migration timescales, as labeled in the inset between the two panels. The simulation setup is as follows: an exterior test particle, initialized on a circular, planar $a'=0.15\,$AU orbit around a $\M=1M_{\odot}$ star, migrates convergently towards a $m=1\times10^{-5}\,M_{\odot}\approx 3\,M_{\oplus}$ planet residing at $a=0.1\,$AU, eventually encountering the 3:2 mean-motion commensurability. Eccentricity damping is applied with a characteristic timescale that is a factor of $\K=1000$ shorter than the migration time, $\tauap$. Simultaneous migration and circularization of the outer orbit is indicated with shaded arrows on the diagram. As $\tauap$ is reduced from $80\,$kyr toward the analytically-predicted critical value of $5\,$kyr (equation \ref{tauacrit}), the equilibrium value of the resonant angle, $\phi$, tends from $\pi$ to $3\pi/2$. The equilibrium value of the eccentricity, on the other hand, stabilizes at $e'=1/\sqrt{4\,\K}\approx 0.016$ in all cases. The behavior of this heavily damped system is fully consistent with expectations provided by analytic theory.}
\label{F:K1000}
\end{figure} 

Accounting for dissipative effects, the equilibrium equations take the form:
\begin{align}
\frac{d\phi}{dt} &= \frac{\partial\Ham}{\partial\Phi}=k\,n'-(k-1)\,n-\frac{\mathcal{F}\,\cos((\phi)_{\rm{eq}})}{\sqrt{2\,(\Phi)_{\rm{eq}}}} \nonumber \\
&-\frac{3\,k\,n'\,((\Psi)_{\rm{eq}}+k\,(\Phi)_{\rm{eq}})}{[\mathcal{L}]}=0 \nonumber \\
\frac{d\Phi}{dt} &= - \frac{\partial\Ham}{\partial\phi}-2\,\K\,\frac{(\Phi)_{\rm{eq}}}{\tauap}=-\mathcal{F}\,\sqrt{2(\Phi)_{\rm{eq}}}\,\sin((\phi)_{\rm{eq}})  \nonumber \\
&-2\,\K\,\frac{(\Phi)_{\rm{eq}}}{\tauap} = 0.
\label{Hammyeq}
\end{align}
The first $(\dot{\phi}=0)$ equation is trivially satisfied. The second $(\dot{\Phi}=0)$ equation, on the other hand, yields the criterion for the shortest migration timescale that allows for resonant capture:
\begin{align}
\tauap &= \frac{\M}{f\,m}\sqrt{\frac{\K}{2\,(k-1)} \frac{a'^3}{\G\,\M}} \nonumber \\ 
&\approx \frac{5}{4}\frac{\M}{m}\sqrt{\frac{\K}{2\,(k-1)^3}}\frac{1}{n},
\label{tauacrit}
\end{align}
where we have used the compact $k\approx k-1$ approximation to evaluate $f\approx4\,k/5$. The expression agrees with the one recently obtained by \citet{2023arXiv230203070H}, who arrived at it through a somewhat distinct -- but ultimately equivalent -- approach.

\subsection{Numerical Experiments}

Whether the process of resonant capture is controlled by stability -- and thus follows the criterion given by equation (\ref{tauacrit}) -- or adiabaticity (as discussed in e.g., \citealt{B15}), depends on the efficiency of orbital circularization. In the limit of weak damping (``small $\K$"), we may reasonably expect that adiabaticity will serve as the more stringent constraint, while stability will dominate in the regime of rapid (``large $\K$") damping. Moreover, the transitionary value of $\K$ is likely to significantly exceed unity, since the eccentricity-damping timescale should be contrasted against the resonance-crossing time -- a quantity that is proportional to $\tauap$, but is much smaller in magnitude. In the remainder of this section, we will use numerical experiments to verify the validity of equation (\ref{tauacrit}), as well as to explore the capture process in the heavily ($\K=1000$) and moderately ($\K=100$) damped regimes.

Our numerical experiments follow the conventional scheme of implementing dissipative effects into an $N-$body framework using the formulae of \citet{PL2000}. For definitiveness, we adopted a physical setup that is reminiscent of short-period extrasolar Super-Earth systems such as Kepler-59 and Kepler-128: a $1\,M_{\odot}$ star encircled by a $m=10^{-5}\,M_{\odot}\approx3\,M_{\oplus}$ planet on a circular orbit at $a=0.1\,$AU \citep{2016ApJ...828...44H, 2020MNRAS.491.5238S}. In addition, we initialized an exterior test-particle into the simulation on a $e'=0$, $a=0.15\,$AU orbit, such that the first commensurability encountered by the system is the 3:2 period-ratio. To drive convergent migration, non-gravitational forces were only applied to the test-particle. The system of ODEs was integrated using the Bulirsch-Stoer algorithm with an accuracy parameter of $\hat{\epsilon}=10^{-10}$.

\begin{figure}[t!]
\centering
\includegraphics[width=\columnwidth]{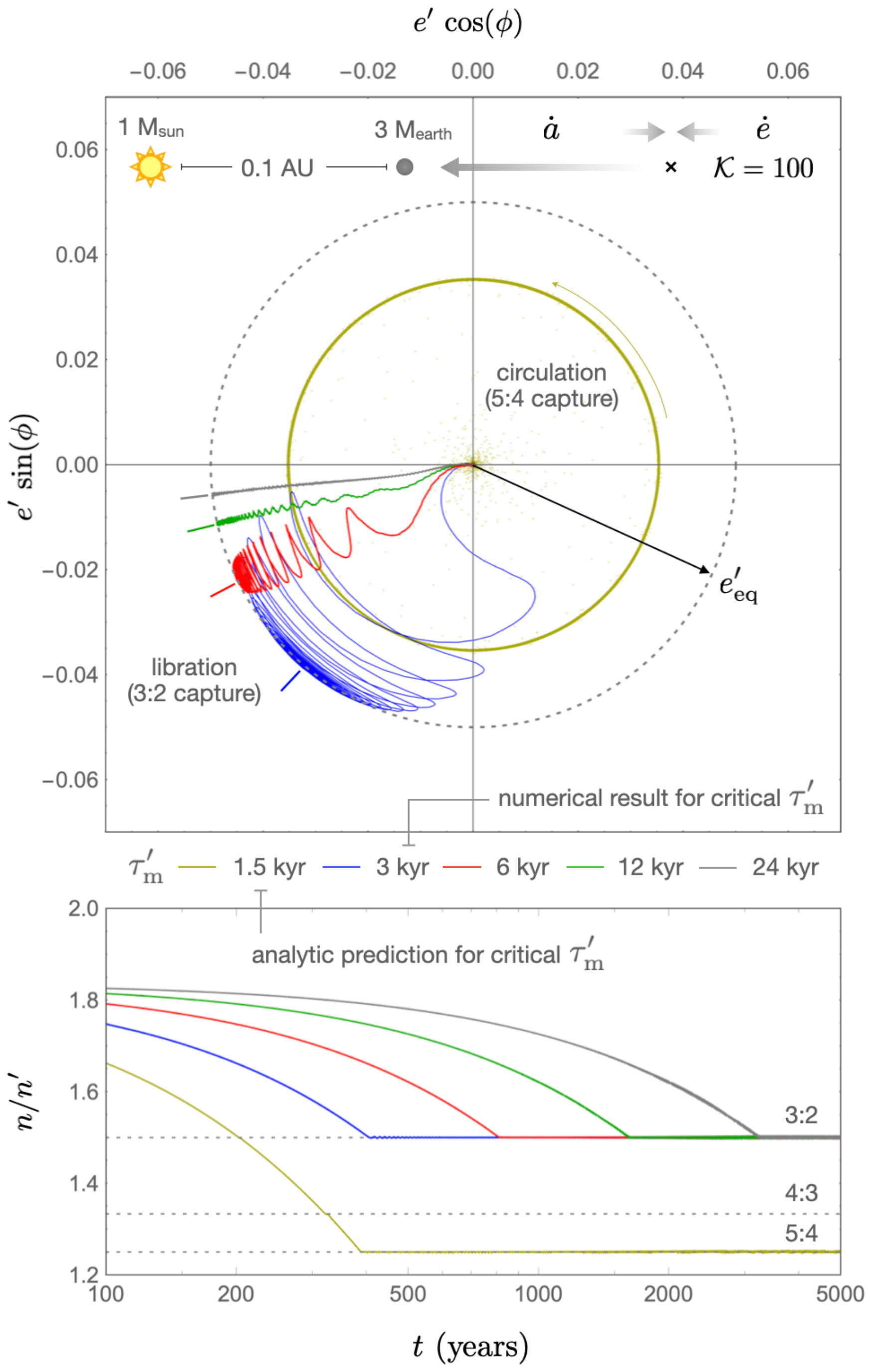}
\caption{Same as Figure \ref{F:K1000} but with a reduced eccentricity damping factor of $\K=100$. While resonant capture ensues for long migration timescales, the analytically-predicted critical value of $\tauap=1.5\,$kyr leads to passage through the 3:2 commensurability and capture into the 5:4 resonance instead. Failure of the stability criterion (\ref{tauacrit}) in this example indicates that the orbital circularization is sufficiently slow that a different mechanism -- namely, adiabaticity -- regulates resonance capture.}
\label{F:K100}
\end{figure}

In both the $\K=1000$ and $\K=100$ simulation suites, we carried out five numerical experiments, first setting $\tau'_{\rm{m}}$ equal to the value given by the stability criterion (\ref{tauacrit}) and doubling it in every run. Figures (\ref{F:K1000}) and (\ref{F:K100}) depict the results of these simulations: the bottom panels show the time-series of the orbital frequency ratio (equal to the period ratio), and the top panels show the phase-space evolution of the system. 

Overall, the results of the numerical experiments with $\K=1000$ conform to the theoretical expectations outlined above. That is, as the test-particle approaches the 3:2 resonance, $e'$ stabilizes at the equilibrium value given by equation (\ref{eeq}) -- shown on the phase-space plot with a gray circle -- independent of the adopted $\tauap$. Conversely, the stationary value of the critical angle ratchets up towards $\phi=3\,\pi/2$ as $\tauap$ approaches the critical value of $\sim5000$ years, as dictated by equation (\ref{tauacrit}). Though all of the shown runs result in stable capture within a 3:2 resonance, we have also confirmed that resonant locking fails for shorter (e.g., $4500$ year) migration timescales, in agreement with the predictions of the analytical theory.

Unlike their more heavily-damped counterparts, the $\K=100$ simulations deviate notably from our analytical stability arguments. While the system equilibrates at the predicted state as long as the convergence time is long, the process of resonance capture is accompanied by a growing libration amplitude, as $\tauap$ approaches the critical value of $\sim1500$ years (recall that it scales as $\propto\sqrt{\K}$). For this reason, the resonant equilibrium becomes compromised at a migration timescale that is a factor of $\sim2$ longer than that predicted by equation (\ref{tauacrit}). Indeed, the run with $\tauap=1500$ years results in passage through the 3:2 commensurability and subsequent capture into the 5:4 resonance (a configuration reminiscent of the Kepler-307 system; \citealt{2016ApJ...820...39J}). Evidently, for the given mass-ratio $m/\M$, a value of $\K$ significantly in excess of $100$ is required to fully suppress the growth of the phase-space area during orbital convergence, giving way to adiabaticity as the process that largely determines the outcome of resonant encounters.

\section{The Unrestricted Problem} \label{sec:section3}

\subsection{Analytical Theory}
With the qualitative picture outlined within the simplified framework of the restricted problem above, generalization of our results to the full resonant three-body problem is relatively undemanding. Retaining the nearly coplanar and low-eccentricity approximations but putting no limits on the planetary mass-ratio $m/m'$, the governing Hamiltonian takes the form:
\begin{align}
\Ham &= -\frac{m^3}{2}\bigg( \frac{\G\,\M}{\Lambda} \bigg)^2 -\frac{m'^3}{2}\bigg( \frac{\G\,\M}{\Lambda'} \bigg)^2 \nonumber \\
&-\mathcal{A}\sqrt{2\,\Gamma}\,\cos(k\,\lambda'-(k-1)\,\lambda+\gamma) \nonumber \\
&-\mathcal{B}\sqrt{2\,\Gamma'}\,\cos(k\,\lambda'-(k-1)\,\lambda+\gamma'),
\label{Hamfull}
\end{align}
where $\mathcal{A} = (\G^2\,\M\,m\,m'^3/\Lambda'^2)\,(g/\sqrt{\Lambda})$ and $\mathcal{B} =(\G^2\,\M\,m\,m'^3/\Lambda'^2)\,(f/\sqrt{\Lambda'})$ are pre-factors similar to $\mathcal{F}$ \citep{BM13,2019AJ....158..238H}. Furthermore, in the above expression, we have used the conventional system of Poincar\'e action-angle variables, defined in the $m\ll\M$ and $e\ll 1$ limit as:
\begin{align}
&\Lambda = m \,\sqrt{\G\,\M\,a} &\lambda = \mathcal{M}+\varpi \nonumber \\
&\Gamma = \Lambda \, e^2/2 & \gamma = - \varpi,
\label{PoincareVars}
\end{align}
with equivalent definitions for the primed quantities.

\begin{figure*}
\centering
\includegraphics[width=\textwidth]{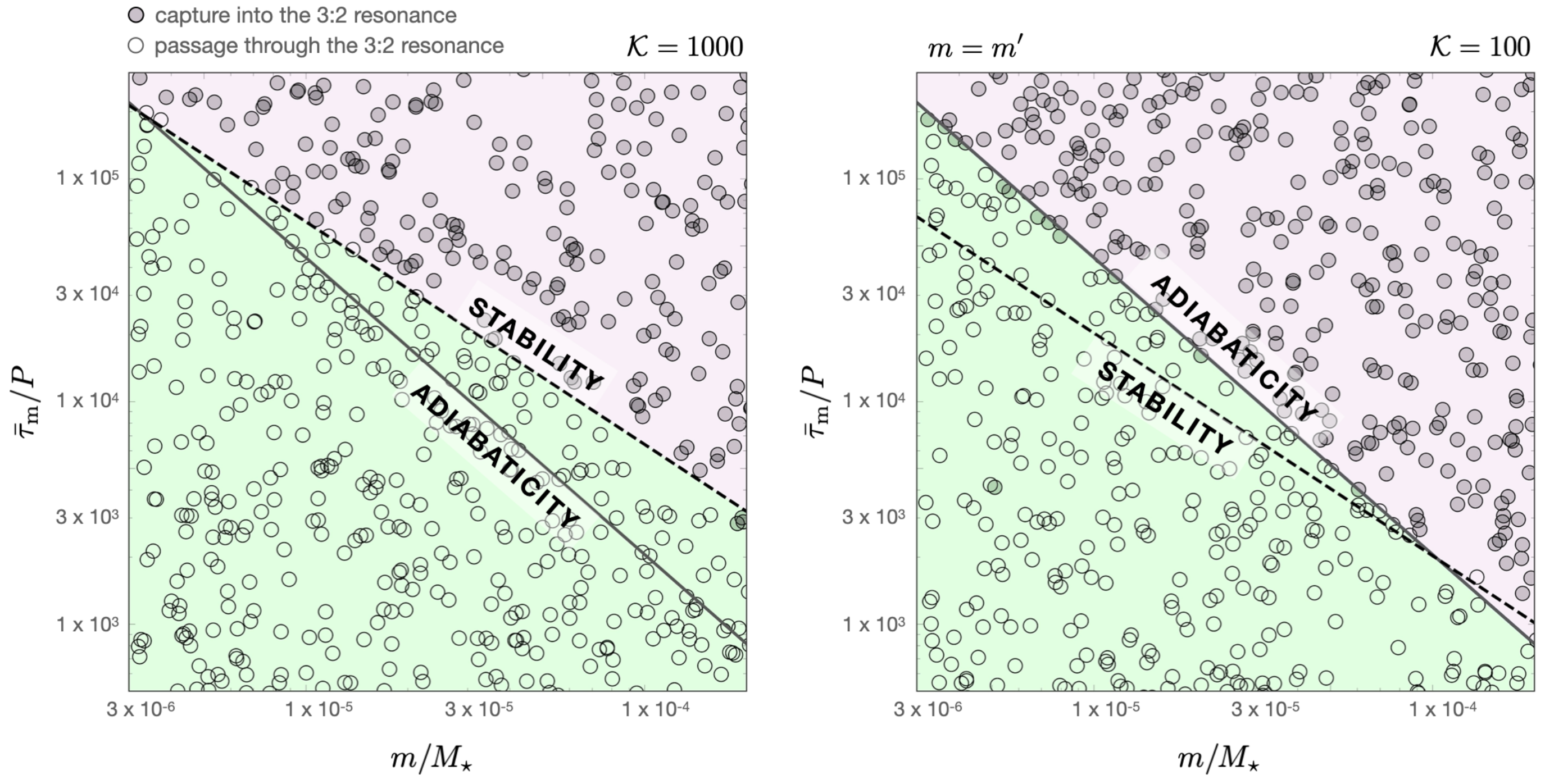}
\caption{A dimensionless mass vs. migration time map of $k=3$ resonance capture for an equal-mass ($m=m'$) planetary system. The filled circles indicate parameter combinations where numerical simulations yield successful resonant capture, whereas empty circles indicate passage through the 3:2 resonance. The left and right panels correspond to the heavily ($\K=1000$) and moderately ($\K=100$) damped regimes, respectively. Both panels additionally show analytic stability and adiabaticity capture criteria as dashed and solid lines. While the stability criterion regulates capture in the heavily-dissipated case, adiabaticity controls the outcome of resonant encounters in the moderately damped regime.}
\label{F:32}
\end{figure*} 

Notice that unlike equation (\ref{Hammy}), the Hamiltonian (\ref{Hamfull}) now contains two harmonic terms, and we have maintained the Keplerian contributions in their unexpanded form (this will not affect our analysis). As in the preceding section, we must now consider the stability of the global fixed point of the resonant Hamiltonian, in presence of migration and dissipation. In fact, similar analyses have previously been carried out in the literature \citep{BM13diss,2019MNRAS.482..530T}, but typically in the limit of weak friction. 

Central to our calculation is the long-term evolution of the semi-major axes, which -- unlike the case of the restricted problem -- can be directed inward, outward, or can be stationary, depending on the signs and magnitudes of the individual planetary migration timescales, $\taua$ and $\tauap$. Nevertheless, the maintenance of the resonant relationship between the orbits necessitates that $\dot{\Lambda}/\Lambda = \dot{\Lambda'}/\Lambda'$. This equality can be computed directly from Hamilton's equation:
\begin{align}
\frac{1}{\Lambda}\frac{d\Lambda}{dt} = -\frac{\partial\,\Ham}{\partial\,\lambda}-\frac{\Lambda\, \big(\taue+2\,\taua (2\,\Gamma/\Lambda)\big) }{2\,\taua\,\taue},
\label{LambdaEOM}
\end{align}
where we have expressed the dissipative contribution (given by equation \ref{taua}) in canonical coordinates. Collecting the Hamiltonian terms on the LHS, we have:
\begin{align}
&\frac{2\,\big(k\,(\Lambda+\Lambda')-\Lambda' \big)}{\Lambda\,\Lambda'}\bigg( \mathcal{A}\,\sqrt{2\,\Gamma}\,\sin(\varphi) + \mathcal{B}\,\sqrt{2\,\Gamma'}\,\sin(\phi) \bigg) \nonumber \\
&=\frac{1}{\taua}+\frac{4\,\Gamma}{\Lambda\,\taue}-\frac{1}{\tauap}-\frac{4\,\Gamma'}{\Lambda'\,\tauep}.
\label{EQN1}
\end{align}
Similarly to equation (\ref{vars}), in the above expression, we have used $\varphi$ and $\phi$ to denote the resonant harmonics containing $\varpi$ and $\varpi'$, respectively.

The dependence of equation (\ref{EQN1}) on the resonant angles as well as the actions $\Gamma$ and $\Gamma'$ can be eliminated by considering the equilibrium of the eccentricities. This equilibrium is given by $\dot{\Gamma}=-\partial\Ham/\partial\gamma-2\,\Gamma/\taue=0$, with an identical expression for $\Gamma'$. In particular, we obtain:
\begin{align}
&\mathcal{A}\,\sqrt{2\,\Gamma}\,\sin(\varphi) = -2\,\Gamma/\taue  \nonumber \\
&\mathcal{B}\,\sqrt{2\,\Gamma'}\,\sin(\phi) = -2\,\Gamma'/\tauep.
\label{eccequilib}
\end{align}
From this expression, it is easy to see how the sinusoidal terms in equation (\ref{EQN1}) can be eliminated. Moreover, in analogy with the preceding section, criticality is attained as $\varphi\rightarrow\pi/2$, $\phi\rightarrow3\pi/2$, which yields $\Gamma = \mathcal{A}^2\,\taue^2/2; \Gamma' = \mathcal{B}^2\,\taue'^2/2$. Upon direct substitution into equation (\ref{EQN1}) and setting $a = ((k-1)/k)^{2/3} a' = \alpha\,a'$, we obtain the criterion for resonant capture:
\begin{align}
&\frac{1}{\tauap} - \frac{1}{\taua} = \frac{2\,k\,\G\,\M}{a'^3}\bigg(\frac{m}{\M}+\frac{m'}{\sqrt{\alpha}\,\M}\bigg) \nonumber \\
&\times \bigg(\frac{g^2\,m'}{\sqrt{\alpha}\,\M}\taue+\frac{k-1}{k}\frac{f^2\,m}{\M}\tauep \bigg)  \nonumber \\
&\approx \frac{32\,\G\,k^3\,\big(m+m'\big)\,\big(m'\,\taue+m\,\tauep\big)}{25\,\M\,a'^3}
\label{citerionfull}
\end{align}
It is trivial to check that this criterion reproduces equation (\ref{tauacrit}) in the limit where $m'\rightarrow0$.

Although here we have derived criterion (\ref{citerionfull}) directly from Hamiltonian (\ref{Hamfull}), an equivalent -- albeit somewhat more mathematically involved -- approach would have been to first reduce $\Ham$ to an integrable form that only contains a single resonant harmonic (see e.g., \citealt{BM13}), and then analyze the stability of its equilibrium under dissipation. This approach yields identical results to those delineated above and is reproduced in the Appendix.

\subsection{Numerical Experiments}
As a quantitative test of the generalized criterion derived above, we have repeated the numerical simulations described in the previous section, this time setting $m'=m$, and sampling a broad range of orbital convergence times and planetary masses, in order to map the capture criterion for the 3:2 resonance on the ($m/\M,\taub/P$) plane, where $P$ is the orbital period of the inner object. For definitiveness, in these simulations, we assigned a common value of $\K$ to both planets (such that $\taue=\tauep=\taub/\K$), but applied orbital decay only to the outer body such that $\taub = \taua\,\tauap/(\taua-\tauap) = \tauap$. Finally, to prevent the system from spiraling onto the central star without careful modeling of the disk's inner edge (e.g., \citealt{Izidoro2017,Izidoro2021}), we simply rescaled both of the semi-major axes at every time-step\footnote{We have simulated the assembly of the Galilean moons into the Laplace resonance using an identical method in a previous study \citep{2020ApJ...894..143B}.}, maintaining the inner planet at $a=0.1\,$AU.

The results of these numerical experiments are shown in Figure (\ref{F:32}). Instances where capture into the 3:2 resonance was successful are shown with filled gray points whereas runs that resulted in passage through the commensurability are shown with empty circles. In both the $\K=1000$ (left panel) and $\K=100$ (right panel) simulation suites, a clear power-law threshold emerges on the diagrams, though the slope of this boundary is subtly distinct. As already shown in the proceeding section, for the adopted mass-ratios, the $\K=1000$ case is regulated by stability, whereas the $\K=100$ case is controlled by adiabaticity. To confirm this expectation, we have over-plotted the $k=3$, $m=m'$ stability and adiabaticity criteria (see \citealt{B15}):
\begin{align}
&\bigg(\frac{\taub}{P}\bigg)_{\rm{stab}}=\frac{5}{32\,\pi}\frac{\M}{m}\sqrt{\frac{\K}{6}} \nonumber \\
&\bigg(\frac{\taub}{P}\bigg)_{\rm{ad}}=\frac{5}{384}\bigg(\frac{125}{4} \bigg)^{1/9}\bigg(\frac{\M}{m}\bigg)^{4/3},
\label{k3criteria}
\end{align}
 with dashed and solid lines, respectively. These two criteria adequately explain the numerical results and highlight the distinct regimes of resonance capture that can ensue within moderately and heavily dissipative planet-formation environments.

\subsection{Scaling for Type-I Migration} \label{sec:typeI}

As a final theme of this section, let us move away from parameterized simulations considered above, and compare our results with more realistic simulations of resonance capture driven by type-I disk migration. For a planet of mass $m$ residing on an orbit with semi-major axis $a$, the characteristic timescale associated with type-I migration is the spiral density wave propagation time \citep{Tanaka2004}:
\begin{align}
\twave = \frac{1}{n}\frac{\M}{m}\frac{\M}{\Sigma\,a^2} \bigg(\frac{h}{r} \bigg)^4,
\end{align}
where $\Sigma$ is the surface density of the nebula, evaluated at the planetary semi-major axis. 

For nearly circular and planar orbits, the eccentricity damping timescale differs from $\twave$ only by a numerical factor of order unity, $f_e = 0.78$, such that $\taue = \twave/f_e$. The semi-major axis damping time, on the other hand, exceeds the wave propagation time by a large margin: $\taua = \twave\,(r/h)^{2}/f_a$. Notably, the proportionality constant depends on the index of the disk's surface density profile, $s$, in a linear manner: $f_a = 2.7 + 1.1\,s$ \citep{Tanaka2002}. Employing these dependencies, \citet{K23} used the surface density of the disk as a physical proxy for migration speed, and formulated the results of their numerical simulation suite in terms of a critical value of $\Sigma$ required for resonance capture.

Carrying out their numerical experiments in a non-evolving, flared, $s=1$ nebula, \citet{K23} simulated the orbital convergence of a pair of equal-mass planets, suppressing disk-driven migration of the inner object, which was initialized at $0.1\,$AU. Intriguingly, they found that the critical value of $\Sigma$ exhibits a clear dependence on the cube of the disk's aspect ratio but is independent of the planetary mass (i.e., critical $\Sigma\propto\,m^0\,h^3$). Let us examine if these scalings can be derived from our analytical criterion.

\begin{figure}[t]
\centering
\includegraphics[width=\columnwidth]{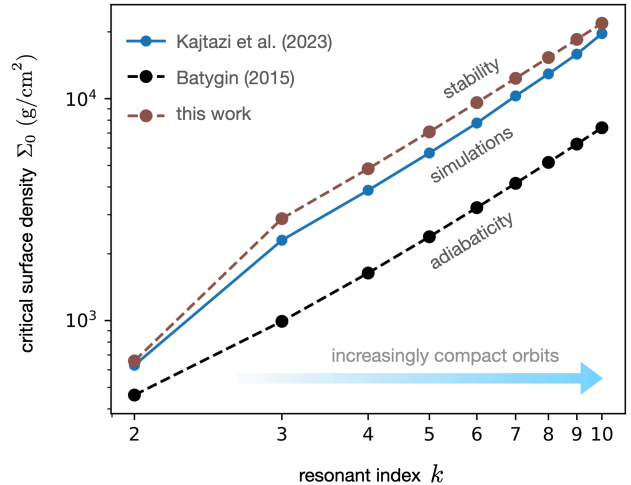}
\caption{Comparison between analytic and numeric determination of resonance capture. Using the nebular surface density as a proxy for the critical migration rate, the results of the \citet{K23} simulation suite ($\K\approx570$) are shown with blue points for a range of resonant indexes, $k$. Analytic stability and adiabaticity criteria are over-plotted with gray and black points, respectively. Although not exact, the critical value of the surface density given by the stability criterion (equation \ref{citerionfull}) matches the numerical data to within $20\%$.}
\label{F:Numsim}
\end{figure}

Plugging in the aforementioned type-I expressions for $\taua$ and $\taue$ into equation (\ref{citerionfull}), we obtain:
\begin{align}
\frac{f_a\,m'\,n'\,\Sigma\,a'^2}{\M}\bigg(\frac{h}{r} \bigg)^{-2}=\frac{128\,\G\,k^3\,m'\,\M}{25\,f_e\,n'\,\Sigma\,a'^5} \bigg(\frac{h}{r} \bigg)^{4},
\end{align}
where we have set $\taub=\tauap;\, m=m'$ following \citet{K23}, and have assumed the compact approximation valid for $k\gtrsim3$ for simplicity. Solving for $\Sigma$ and relating the value to the reference surface density, $\Sigma_0$, at $r_0=1\,$AU, we recover the scalings found in simulations:
\begin{align}
\frac{\Sigma\,a'^2}{\M}= \frac{\Sigma_0\,r_0\,a'}{\M} =\sqrt{\frac{2\,k^3}{f_a\,f_e}}\bigg(\frac{h}{r}\bigg)^3.
\end{align}
As a concluding step, to assess the degree of quantitative agreement between theory and numerical experiments, in Figure (\ref{F:Numsim}), we plot our analytical criterion for a range of resonance indexes together with the numerical results of \cite[][their Fig. 6]{K23}. Though the agreement is not exact, we  find that the analytical result does not deviate from the numerical findings by more than 20\%.

\section{Discussion} \label{sec:section4}

Capture of planets into mean-motion resonances is an expected outcome of evolution within protoplanetary nebulae, and in this work, we have derived an analytic criterion for resonance locking in presence of migration and orbital circularization. Our theoretical framework is based upon a perturbative treatment of the unrestricted gravitational three-body problem (\citealt{Peale1986,MD99} and the refs therein), and thus assumes small eccentricities and inclinations, while placing no restrictions on the planet-planet mass ratio. Fundamentally, our derivation rests upon the stability analysis of the resonant equilibrium in presence of dissipation, and complements the previously obtained criterion that stems from a consideration of adiabaticity \citep{B15}.

We note that the stability argument considered here concerns the elementary question of the existence of an equilibrium point for the resonant variables and not the long-term post-capture evolution. To this end, it has been shown by \citet{Goldreich2014,Deck15,2018MNRAS.481.1538X} that depending on the ratio of the planetary masses and eccentricity damping timescales, systems where resonant capture is initially successful can eventually escape from resonance by way of over-stability. Although beyond the scope of our study, these dynamics, along with stochastic growth of the libration amplitude by nebular turbulence \citep{2008ApJ...683.1117A,2017AJ....153..120B}, synodic modulation of resonant angles \citep{2020MNRAS.494.4950P,GB22b}, etc., give rise to additional constraints on the kind of architectures that can be established within protoplanetary disks. 

It is interesting to note that although the criteria for resonant capture based upon stability and adiabaticity yield distinct scalings with the planetary mass, for typical parameters pertinent to planetary migration within circumstellar disks, they can give results that are quantitatively similar. Ultimately, for resonant locking to ensue, both criteria -- stability and adiabaticity -- must be satisfied, such that whichever one yields the longer critical timescale for orbital convergence plays the controlling role. Accordingly, the specific capture regime is dictated by the combination of the ratio of the cumulative planetary mass to the mass of the star as well as the ratio of the migration and circularization timescales. 

For sub-Jovian planet pairs, the analysis carried out in section (\ref{sec:section3}) indicates that the outcomes of resonant encounters in systems with $\K=10^2$ and $\K=10^3$ are determined by adiabaticity and stability respectively, such that the transitionary value of the timescale ratio lies in between these two regimes. While the appropriate value of $\K$ is bound to be case-specific, comparison of our results with the published simulation suite of \citet{K23}, indicates the that stability criterion (\ref{citerionfull}) reproduces the scalings seen in the numerical experiments. Still, it remains important to keep in mind that within the context of the flared disk model of \citet{K23}, the disk aspect ratio at a stellocentric distance of $0.1\,$AU (where the resonant encounter is set up to take place) is $(h/r) = 0.033 \times (0.1)^{1/4}\approx0.019$, corresponding to $\K\approx 570$. Thus, it is reasonable to expect that if the resonant encounters were to instead take place at $10\,$AU, where $(h/r)\gtrsim0.05$ and $\K\lesssim100$, adiabaticity would have played the deterministic role. In summary, the analytical framework developed here adds to our understanding of the dynamics of resonant encounters, and serves as a basis for further interpretation of the early evolution of planetary systems.

\paragraph{Acknowledgments.} K. B. is grateful to Caltech, the David and Lucile Packard Foundation, and the National Science Foundation (grant number: AST 2109276) for their generous support. During the preparation of this paper, we have become aware that Huang \& Ormel (2023, submitted) arrived at similar arguments simultaneously and independently.

\begin{appendix} \label{append}

\section*{Derivation of the Stability Criterion From an Integrable Hamiltonian}
\label{app:gencrit}
In this appendix, we present an alternative derivation of the stability criterion given by equation (\ref{citerionfull}), based upon the reduction of the resonant three-body problem to an integrable Hamiltonian. In particular, we follow the formalism outlined in \cite{Deck15}, switching to their notation for consistency. Because this reduction is well documented \citep[\emph{e.g.}][]{BM13,2013ApJ...774..129D,2017A&A...607A..35P,2019AJ....158..238H}, we limit ourselves to recalling the relevant variables and refer the interested reader to the cited works.
	
As in the main text, we consider the general problem of two planets of mass $m_1$ and $m_2$ orbiting a star of mass $M_\star$ in the plane. We assume that the planets experience a convergent migration and are close to the crossing of the $k$:$k-1$ resonance. The integrable Hamiltonian has the form:
	\begin{align}
	\Ham = -\frac{1}{2}(\Phi'-\Gamma')^2 -\sqrt{2\Phi'}\cos(\phi),
	\end{align}
where $\Phi'$ is the renormalized Sessin variable \citep{1984CeMec..32..307S}
	\begin{align}
		\Phi' &= \frac{1}{Q} \frac{k}{k-1}\frac{\ares}{\zeta+\ares}\frac{\zeta\sqrt{\ares}}{2(R^2+\zeta\sqrt{\ares})}\sigma^2\label{eq:phip},\\
		\sigma^2 &= R^2e_1^2+e_2^2-2Re_1e_2\cos(\Delta\varpi),\nonumber
	\end{align}
	where $R = |f_{k,27}(\ares)|/f'_{k,31}(\ares)$, $\zeta = m_1/m_2$, $\ares=(1-1/k)^{2/3}$ and the functions $f_{27/31}$ are the resonant coefficients \citep{MD99}.
	$Q$ is a renormalization factor for the actions
	\begin{equation}
		Q = \varepsilon_{\rm p}^{2/3}\zeta\frac{\ares^{5/6}}{k-1}\left(\frac{f'^2_{k,31}}{9k(1+\zeta)^2}\frac{R^2+\zeta\sqrt{\ares}}{(\ares+\zeta)^5}\right)^{1/3}
	\end{equation}
	defined by \cite[][their Appendix A]{Deck15} and $\varepsilon_{\rm p} = (m_1+m_2)/M_\star$.
	The angle $\phi$ is conjugated with $\Phi'$ and is the generalized resonant angle.
	$\Gamma'$, on the other hand, is a parameter that is related to the system's angular momentum.
	
	Maintaining the notation of \cite{Deck15}, we have the relative migration timescale
	\begin{equation}
		\frac{1}{\tau_a} = \frac{1}{\tau_{a,2}}-\frac{1}{\tau_{a,1}},
	\end{equation}
	and the eccentricity damping timescale
	\begin{equation}
		\frac{1}{\tau_e} = \frac{1}{\tau_{e,1}}+\frac{\zeta }{\tau_{e,2}}.
		\label{eq:DBtaue}
	\end{equation}
	Note that without making any approximations in the computation of the dissipation onto the action $\Phi'$, one obtains the following eccentricity damping timescale
	\begin{equation}
		\frac{1}{\tau_e} = \frac{1}{\tau_{e,1}}+\frac{\sqrt{\alpha}}{R^2}\frac{\zeta }{\tau_{e,2}}
		\label{eq:bettertaue}
	\end{equation}
	This expression slightly differs from that given in \citet{Deck15} by a few percent for \(k\geq2\) but is more accurate for the 2:1 MMR. \cite{Deck15} also define the timescale
	\begin{equation}
		\frac{1}{\tau_{a,e}} = \frac{1}{\tau_{e,1}}-\frac{{\alpha}}{R^2}\frac{\zeta^2 }{\tau_{e,2}}.
	\end{equation}
	
	The damping on the action $\Phi'$ is expressed as
	\begin{equation}
		\left.\deriv{\Phi'}{t'}\right|_{\rm dis} = C_0 \Phi'
	\end{equation}
	where we assume $\tau_a\gg\tau_e$; neglecting higher order terms in eccentricities, we have:
	\begin{equation}
		C_0 = -\frac{2\gamma_1}{\tau_e},
	\end{equation}
	where
	\begin{equation}
		\gamma_1 = \frac{R^2}{R^2+\zeta\sqrt{\alpha}},
	\end{equation}
	and for completness we define $\gamma_2 = 1-\gamma_1$.
	Damping on the Hamiltonian parameter $\Gamma'$ is expressed as
	\begin{equation}
		\left.\deriv{\Gamma'}{t'}\right|_{\rm dis} = A_0+ A_1 \Phi'
	\end{equation}
	with 
	\begin{equation}
		A_0 = \frac{\zeta\sqrt{\alpha}}{2Qk\eta_2^2}\frac{1}{\tau_a}\quad \mathrm{and}\quad
		A_1 = \frac{-2p\gamma_1}{k\eta_2}\frac{1}{\tau_{a,e}},
	\end{equation}
	where
	\begin{equation}
		\eta_2 = \frac{k-1}{k}+\zeta\sqrt{\alpha}
	\end{equation}
	with $\eta_1 = \eta_2/(\zeta\sqrt{\alpha})$.
	
	The equations of motion\footnote{Expressed in terms of the renormalized time $t'$.} of the dissipative problems are
	\begin{align}
		\deriv{\Phi'}{t'} &= -\sqrt{2\Phi'}\sin\phi +\frac{\eta_2^3 C_0\Phi'}{n_2Q|\mathcal{K}_2|}\\
		\deriv{\phi}{t'} &= -(\Phi'-\Gamma')-\frac{1}{\sqrt{2\Phi'}}\cos\phi\\
		\deriv{\Gamma'}{t'} &= \frac{\eta_2^3}{n_2Q|\mathcal{K}_2|} (A_0+(A_1+C_0)\Phi')
	\end{align}
	where $n_j$ is the mean motion and $\mathcal{K}_2 = -3(k-1)^2(\ares+\zeta)^5/(\zeta\ares)$  is the second order derivative of the Keplerian Hamiltonian at the exact resonance.
	Looking for an equilibrium, we have
	\begin{equation}
		\Phi_{\rm eq}' = \frac{-A_0}{C_0+A_1}
	\end{equation}
	and,
	\begin{equation}
		\sin \phi_{\rm eq} = \frac{\eta_2^3\sqrt{\Phi_{\rm eq}'}C_0}{\sqrt{2}Q| \mathcal{K}_2|n_2}.
		\label{eq:sinphieq}
	\end{equation}
	
	The equilibrium exists if the absolute value of the right hand side of the above expression is smaller than 1.
	Accordingly, one can derive a condition on the migration speed $\tau_a$ that is, without further approximation:
	\begin{equation}
		\frac{1}{\tau_an_2}<\sqrt{2}\varepsilon_{\rm p}\sqrt{\frac{\tau_e}{\tau_a}}G(k,\zeta,p_{a,e})
		\label{eq:critgen}
	\end{equation}
	where $G$ is a function solely of the resonant index $k$, the planet mass ratio $\zeta$ and the ratio of the eccentricity damping timescales through the parameter $p_{a,e}$
	\begin{equation}
		G(k,\zeta,p_{a,e}) = \frac{\zeta\sqrt{\ares}+R^2}{1+\zeta}\sqrt{\frac{\frac{k-1}{k}+\zeta\sqrt{\ares}}{R^2\ares}}f'_{k,31}\sqrt{k}\sqrt{1+p_{a,e}},
	\end{equation}
	where
	\begin{equation}
		p_{a,e} = \frac{p}{k-1+k\zeta\sqrt{\ares}}\frac{\tau_e}{\tau_{a,e}}.
	\end{equation}
	The expression \eqref{eq:critgen} is equivalent to Eq.~\eqref{citerionfull} for $p=0$.
	
	In the case of an outer test particle, $\zeta\to +\infty$, using expression Eq.~\eqref{eq:bettertaue} for $\tau_e$, the criterion becomes
	\begin{equation}
		\frac{1}{\tau_an_2}<\sqrt{2}\varepsilon_{\rm p}\sqrt{\frac{\tau_{e,2}}{\tau_a}}f'_{k,31}\sqrt{k}\sqrt{1-\frac{p}{k}}.
	\end{equation}
	Taking $p=0$ yields the expression given by equation ~\eqref{tauacrit} derived from the restricted problem. Similarly $p=1$ corresponds to the result obtained by \cite{2023arXiv230203070H}.

\end{appendix}


\begin{thebibliography}{00}


\bibitem[Adams et al.(2008)]{2008ApJ...683.1117A} Adams, F.~C., Laughlin, G., \& Bloch, A.~M.\ 2008, \apj, 683, 1117. doi:10.1086/589986


\bibitem[Batygin \& Brown(2010)]{2010ApJ...716.1323B} Batygin, K. \& Brown, M.~E.\ 2010, \apj, 716, 1323. doi:10.1088/0004-637X/716/2/1323

\bibitem[Batygin \& Morbidelli(2013a)]{BM13} Batygin, K. \& Morbidelli, A.\ 2013a, \aap, 556, A28. doi:10.1051/0004-6361/201220907

\bibitem[Batygin \& Morbidelli(2013b)]{BM13diss} Batygin, K. \& Morbidelli, A.\ 2013b, \aj, 145, 1. doi:10.1088/0004-6256/145/1/1

\bibitem[Batygin(2015)]{B15} Batygin, K.\ 2015, \mnras, 451, 2589. doi:10.1093/mnras/stv1063

\bibitem[Batygin \& Adams(2017)]{2017AJ....153..120B} Batygin, K. \& Adams, F.~C.\ 2017, \aj, 153, 120. doi:10.3847/1538-3881/153/3/120

\bibitem[Batygin \& Morbidelli(2020)]{2020ApJ...894..143B} Batygin, K. \& Morbidelli, A.\ 2020, \apj, 894, 143. doi:10.3847/1538-4357/ab8937

\bibitem[Batygin \& Morbidelli(2023)]{BM23} Batygin, K. \& Morbidelli, A.\ 2023, Nature Astronomy. doi:10.1038/s41550-022-01850-5


\bibitem[Cresswell \& Nelson(2006)]{Cresswell2006} Cresswell, P. \& Nelson, R. P.\ 2006, \aap, 450, A833. doi:10.1051/0004-6361:20054551


\bibitem[Deck et al.(2013)]{2013ApJ...774..129D} Deck, K.~M., Payne, M., \& Holman, M.~J.\ 2013, \apj, 774, 129. doi:10.1088/0004-637X/774/2/129

\bibitem[Deck \& Batygin(2015)]{Deck15} Deck, K.~M. \& Batygin, K.\ 2015, \apj, 810, 119. doi:10.1088/0004-637X/810/2/119

\bibitem[Dai et al.(2023)]{2023AJ....165...33D} Dai, F., Masuda, K., Beard, C., et al.\ 2023, \aj, 165, 33. doi:10.3847/1538-3881/aca327


\bibitem[Dermott(1968a)]{Dermott1968a} Dermott, S.~F.\ 1968a, \mnras, 141, 349. doi:10.1093/mnras/141.3.349

\bibitem[Dermott(1968b)]{Dermott1968b} Dermott, S.~F.\ 1968b, \mnras, 141, 363. doi:10.1093/mnras/141.3.363





\bibitem[Goldberg \& Batygin(2021)]{2021AJ....162...16G} Goldberg, M. \& Batygin, K.\ 2021, \aj, 162, 16. doi:10.3847/1538-3881/abfb78

\bibitem[Goldberg \& Batygin(2022a)]{GB22a} Goldberg, M. \& Batygin, K.\ 2022, \aj, 163, 201. doi:10.3847/1538-3881/ac5961

\bibitem[Goldberg et al.(2022b)]{GB22b} Goldberg, M., Batygin, K., \& Morbidelli, A.\ 2022, \icarus, 388, 115206. doi:10.1016/j.icarus.2022.115206


\bibitem[Goldreich(1965)]{Goldreich1965} Goldreich, P.\ 1965, \mnras, 130, 159. doi:10.1093/mnras/130.3.159

\bibitem[Goldreich \& Tremaine(1980)]{1980ApJ...241..425G} Goldreich, P. \& Tremaine, S.\ 1980, \apj, 241, 425. doi:10.1086/158356

\bibitem[Goldreich \& Schlichting(2014)]{Goldreich2014} Goldreich, P. \& Schlichting, H.~E.\ 2014, \aj, 147, 32. doi:10.1088/0004-6256/147/2/32

\bibitem[Go{\'z}dziewski et al.(2016)]{2016MNRAS.455L.104G} Go{\'z}dziewski, K., Migaszewski, C., Panichi, F., et al.\ 2016, \mnras, 455, L104. doi:10.1093/mnrasl/slv156


\bibitem[Hadden \& Lithwick(2016)]{2016ApJ...828...44H} Hadden, S. \& Lithwick, Y.\ 2016, \apj, 828, 44. doi:10.3847/0004-637X/828/1/44

\bibitem[Hadden(2019)]{2019AJ....158..238H} Hadden, S.\ 2019, \aj, 158, 238. doi:10.3847/1538-3881/ab5287

\bibitem[Henrard(1982)]{1982CeMec..27....3H} Henrard, J.\ 1982, Celestial Mechanics, 27, 3. doi:10.1007/BF01228946

\bibitem[Henrard \& Lemaitre(1983)]{1983CeMec..30..197H} Henrard, J. \& Lemaitre, A.\ 1983, Celestial Mechanics, 30, 197. doi:10.1007/BF01234306

\bibitem[Huang \& Ormel(2023)]{2023arXiv230203070H} Huang, S. \& Ormel, C.\ 2023, arXiv:2302.03070. doi:10.48550/arXiv.2302.03070


\bibitem[Izidoro et al.(2017)]{Izidoro2017} Izidoro, A., Ogihara, M., Raymond, S.~N., et al.\ 2017, \mnras, 470, 1750. doi:10.1093/mnras/stx1232

\bibitem[Izidoro et al.(2021)]{Izidoro2021} Izidoro, A., Bitsch, B., Raymond, S.~N., et al.\ 2021, \aap, 650, A152. doi:10.1051/0004-6361/201935336



\bibitem[Jontof-Hutter et al.(2016)]{2016ApJ...820...39J} Jontof-Hutter, D., Ford, E.~B., Rowe, J.~F., et al.\ 2016, \apj, 820, 39. doi:10.3847/0004-637X/820/1/39


\bibitem[Kajtazi et al.(2023)]{K23} Kajtazi, K., Petit, A.~C., \& Johansen, A.\ 2023, \aap, 669, A44. doi:10.1051/0004-6361/202244460


\bibitem[Lee \& Peale(2002)]{2002ApJ...567..596L} Lee, M.~H. \& Peale, S.~J.\ 2002, \apj, 567, 596. doi:10.1086/338504

\bibitem[Luger et al.(2017)]{2017NatAs...1E.129L} Luger, R., Sestovic, M., Kruse, E., et al.\ 2017, Nature Astronomy, 1, 0129. doi:10.1038/s41550-017-0129


\bibitem[Masset et al.(2006)]{2006ApJ...642..478M} Masset, F.~S., Morbidelli, A., Crida, A., et al.\ 2006, \apj, 642, 478. doi:10.1086/500967

\bibitem[Mestel(1963)]{1963MNRAS.126..553M} Mestel, L.\ 1963, \mnras, 126, 553. doi:10.1093/mnras/126.6.553

\bibitem[Mills et al.(2016)]{2016Natur.533..509M} Mills, S.~M., Fabrycky, D.~C., Migaszewski, C., et al.\ 2016, \nat, 533, 509. doi:10.1038/nature17445

\bibitem[Murray \& Dermott(1999)]{MD99} Murray, C.~D. \& Dermott, S.~F.\ 1999, Solar system dynamics. Cambridge, UK: Cambridge University Press, ISBN 0-521-57295-9



\bibitem[Neishtadt(1984)]{1984PriMM..48..197N} Neishtadt, A.~I.\ 1984, Prikladnaia Matematika i Mekhanika, 48, 197

\bibitem[Nesvorn{\'y} \& Morbidelli(2012)]{2012AJ....144..117N} Nesvorn{\'y}, D. \& Morbidelli, A.\ 2012, \aj, 144, 117. doi:10.1088/0004-6256/144/4/117

\bibitem[Nesvorn{\'y} et al.(2022)]{2022ApJ...925...38N} Nesvorn{\'y}, D., Chrenko, O., \& Flock, M.\ 2022, \apj, 925, 38. doi:10.3847/1538-4357/ac36cd



\bibitem[Papaloizou \& Larwood(2000)]{PL2000} Papaloizou, J.~C.~B. \& Larwood, J.~D.\ 2000, \mnras, 315, 823. doi:10.1046/j.1365-8711.2000.03466.x

\bibitem[Peale(1986)]{Peale1986} Peale, S.~J.\ 1986, IAU Colloq. 77: Some Background about Satellites, 159

\bibitem[Petit et al.(2017)]{2017A&A...607A..35P} Petit, A.~C., Laskar, J., \& Bou{\'e}, G.\ 2017, \aap, 607, A35. doi:10.1051/0004-6361/201731196

\bibitem[Petit et al.(2020)]{2020MNRAS.496.3101P} Petit, A.~C., Petigura, E.~A., Davies, M.~B., et al.\ 2020, \mnras, 496, 3101. doi:10.1093/mnras/staa1736

\bibitem[Pichierri et al.(2018)]{2018CeMDA.130...54P} Pichierri, G., Morbidelli, A., \& Crida, A.\ 2018, Celestial Mechanics and Dynamical Astronomy, 130, 54. doi:10.1007/s10569-018-9848-2

\bibitem[Pichierri et al.(2019)]{2019A&A...625A...7P} Pichierri, G., Batygin, K., \& Morbidelli, A.\ 2019, \aap, 625, A7. doi:10.1051/0004-6361/201935259

\bibitem[Pichierri \& Morbidelli(2020)]{2020MNRAS.494.4950P} Pichierri, G. \& Morbidelli, A.\ 2020, \mnras, 494, 4950. doi:10.1093/mnras/staa1102

\bibitem[Pichierri et al.(2022)]{2022arXiv221203608P} Pichierri, G., Bitsch, B., \& Lega, E.\ 2022, arXiv:2212.03608. doi:10.48550/arXiv.2212.03608




\bibitem[Saad-Olivera et al.(2020)]{2020MNRAS.491.5238S} Saad-Olivera, X., Martinez, C.~F., Costa de Souza, A., et al.\ 2020, \mnras, 491, 5238. doi:10.1093/mnras/stz3369

\bibitem[Sessin \& Ferraz-Mello(1984)]{1984CeMec..32..307S} Sessin, W. \& Ferraz-Mello, S.\ 1984, Celestial Mechanics, 32, 307. doi:10.1007/BF01229087


\bibitem[Tanaka et al.(2002)]{Tanaka2002} Tanaka, H., Takeuchi, T., \& Ward, W.~R.\ 2002, \apj, 565, 1257. doi:10.1086/324713

\bibitem[Tanaka \& Ward(2004)]{Tanaka2004} Tanaka, H. \& Ward, W.~R.\ 2004, \apj, 602, 388. doi:10.1086/380992

\bibitem[Terquem \& Papaloizou(2019)]{2019MNRAS.482..530T} Terquem, C. \& Papaloizou, J.~C.~B.\ 2019, \mnras, 482, 530. doi:10.1093/mnras/sty2693




\bibitem[Ward(1997)]{1997Icar..126..261W} Ward, W.~R.\ 1997, \icarus, 126, 261. doi:10.1006/icar.1996.5647

\bibitem[Wisdom(1986)]{1986CeMec..38..175W} Wisdom, J.\ 1986, Celestial Mechanics, 38, 175. doi:10.1007/BF01230429


\bibitem[Xu et al.(2018)]{2018MNRAS.481.1538X} Xu, W., Lai, D., \& Morbidelli, A.\ 2018, \mnras, 481, 1538. doi:10.1093/mnras/sty2406



\end{thebibliography}
\end{document}